\documentclass[]{aastex631}
\usepackage{amsmath,bm}
\usepackage{lipsum}
\usepackage{svg}
\usepackage{float}

\begin{document}

\title{Rare Event Classification with Weighted Logistic Regression for Identifying Repeating Fast Radio Bursts}

\author[0000-0002-3654-4662]{Antonio Herrera-Martin}
\affiliation{David A.~Dunlap Department of Astronomy \& Astrophysics, University of Toronto, 50 St.~George Street, Toronto, ON M5S 3H4, Canada}
\affiliation{Department of Statistical Science, University of Toronto, Ontario Power Building, 700 University Avenue, 9th Floor, Toronto, ON, Canada M5G 1Z5}

\author[0000-0002-1348-8063]{Radu V. Craiu}
\affiliation{Department of Statistical Science, University of Toronto, Ontario Power Building, 700 University Avenue, 9th Floor, Toronto, ON, Canada M5G 1Z5}
 
\author[0000-0003-3734-8177]{Gwendolyn M. Eadie}
\affiliation{David A.~Dunlap Department of Astronomy \& Astrophysics, University of Toronto, 50 St.~George Street, Toronto, ON M5S 3H4, Canada}
\affiliation{Department of Statistical Science, University of Toronto, Ontario Power Building, 700 University Avenue, 9th Floor, Toronto, ON, Canada M5G 1Z5}

\author[0000-0002-9761-4353]{David C. Stenning}
\affiliation{Department of Statistics and Actuarial Science, Simon Fraser University, 8888 University Drive, Burnaby, B.C., V5A 1S6 Canada}

\author[0000-0001-5628-7256]{Derek Bingham}
\affiliation{Department of Statistics and Actuarial Science, Simon Fraser University, 8888 University Drive, Burnaby, B.C., V5A 1S6 Canada}
% \author[0000-0001-6422-8125]{Amanda Cook}
% \affiliation{David A.~Dunlap Department of Astronomy \& Astrophysics, University of Toronto, 50 St.~George Street, Toronto, ON M5S 3H4, Canada}

\author[0000-0002-3382-9558]{Bryan M. Gaensler}
\affiliation{Department of Astronomy and Astrophysics, University of California Santa Cruz, 1156 High Street, Santa Cruz, CA 95064, USA}
\affiliation{Dunlap Institute for Astronomy \& Astrophysics, University of Toronto, 50 St.~George Street, Toronto, ON M5S 3H4, Canada}
\affiliation{David A.~Dunlap Department of Astronomy \& Astrophysics, University of Toronto, 50 St.~George Street, Toronto, ON M5S 3H4, Canada}

\author[0000-0002-4795-697X]{Ziggy Pleunis}
\affiliation{Dunlap Institute for Astronomy \& Astrophysics, University of Toronto, 50 St.~George Street, Toronto, ON M5S 3H4, Canada}
\affiliation{Anton Pannekoek Institute for Astronomy, University of Amsterdam, Science Park 904, 1098 XH Amsterdam, The Netherlands}
\affiliation{ASTRON, Netherlands Institute for Radio Astronomy, Oude Hoogeveensedijk 4, 7991 PD Dwingeloo, The Netherlands}

\author[0000-0002-7374-7119]{Paul Scholz}
\affiliation{Department of Physics and Astronomy, York University, 4700 Keele Street, Toronto, ON MJ3 1P3, Canada}
\affiliation{Dunlap Institute for Astronomy \& Astrophysics, University of Toronto, 50 St.~George Street, Toronto, ON M5S 3H4, Canada}

\author[0000-0001-7348-6900]{Ryan Mckinven}
\affiliation{Department of Physics, McGill University, 3600 rue University, Montr\'eal, QC H3A 2T8, Canada}
\affiliation{Trottier Space Institute, McGill University, 3550 rue University, Montr\'eal, QC H3A 2A7, Canada}

\author[0009-0008-6166-1095]{Bikash Kharel}
\affiliation{Department of Physics and Astronomy, West Virginia University, P.O. Box 6315, Morgantown, WV 26506, USA }

\author[0000-0002-4279-6946]{Kiyoshi W. Masui}
\affiliation{MIT Kavli Institute for Astrophysics and Space Research, Massachusetts Institute of Technology, 77 Massachusetts Ave, Cambridge, MA
02139, USA}
\affiliation{Department of Physics, Massachusetts Institute of Technology, 77 Massachusetts Ave, Cambridge, MA 02139, USA}
% \collaboration{20}{(AAS Journals Data Editors)}

% \author{C}
% \affiliation{Arizona State University}
% \affiliation{AAS Journals Associate Editor-in-Chief}

% \author{Amy Hendrickson}
% \altaffiliation{AASTeX v6+ programmer}
% \affiliation{TeXnology Inc.}

% \author{Julie Steffen}
% \affiliation{AAS Director of Publishing}
% \affiliation{American Astronomical Society \\
% 1667 K Street NW, Suite 800 \\
% Washington, DC 20006, USA}



\begin{abstract}

An important task in the study of fast radio bursts (FRBs) remains the automatic classification of repeating and non-repeating sources based on their morphological properties. We propose a statistical model that considers a modified  logistic regression to classify FRB sources. The classical logistic regression model is modified to accommodate the small proportion of repeaters in the data, a feature that is likely due to the sampling procedure and duration and is not a characteristic of the population of FRB sources. The weighted logistic regression hinges on the choice of a tuning parameter that represents the true proportion $\tau$ of repeating FRB sources in the entire population.
The proposed method has a sound statistical foundation, direct interpretability, and operates with only 5 parameters, enabling quicker retraining with added data.  
Using the CHIME/FRB Collaboration sample of repeating and non-repeating FRBs and numerical experiments, we achieve a classification accuracy for repeaters of nearly 75\% or higher  when $\tau$ is set in the range of $50$ to $60$\%. This implies a tentative high proportion of repeaters, which is  surprising, but is also in agreement with recent estimates of $\tau$ that are obtained using  other methods.
\end{abstract}

\keywords{}

\section{Introduction} \label{sec:intro}

Fast radio bursts (FRBs) represent an enigmatic phenomenon in astrophysics. FRBs are dispersed, isolated, millisecond-long radio pulses that are similar in appearance to single pulses from Galactic pulsars. The arrival of these radio pulses show a frequency-dependent delay (quantified by the dispersion measure or DM) due to the electromagnetic wave's path through free electrons in the Universe. FRBs have the defining characteristic of a DM that exceeds the maximum DM expected from our Galaxy, suggesting that they are very luminous and of extragalactic origin~\citep{lorimer2007, yaomanchester2017}. 

The first FRB was identified in archival Parkes multibeam pulsar survey data by~\cite{lorimer2007}, and was suggestive of extragalactic origin. Since then, there has been rapid progress in the observation of these enigmatic events and their use as probes of the intergalactic medium~\citep{shami2021, chime_2021,frbreview2022}. The discovery of FRBs has opened up new research avenues in astrophysics as they have the potential to help us better understand the distribution of matter in the universe or the nature of dark matter~\citep{Lin2021, Zhao2023}.

The biggest mystery about FRBs is their origin or progenitor object(s). The FRB enigma is made more mysterious by the fact that some FRBs are observed to burst repeatedly (repeaters), while others have only been observed to burst once (non-repeaters) \citep{Spitler2016}. 
Models that explain the progenitors of FRBs are thus typically assigned to one of two broad categories. The first category considers non-cataclysmic explanations, while the second category assumes FRBs are the result of a catastrophic event that destroys the astrophysical 
source. 
In the early years of FRB discoveries, the lack of repeating FRBs supported catastrophic models.  The assumption of two distinct subpopulations of FRBs (repeaters and non-repeaters) is now supported by arguments made by \cite{Pleunis_2021}, which are based on the morphological differences between the repeating and non-repeating FRBs. While the number of FRB repeater sources continues to grow, the total published sample is currently only 53\footnote{\url{https://www.wis-tns.org/} accessed May 30th, 2024}.

The number of known FRB repeaters is small compared to the total FRB population (roughly 2.6\% are repeaters, \citealt{chime2023}), but observational selection effects strongly influence this number. The CHIME telescope, the world's leading detector of FRBs, relies on the rotation of the Earth to observe the whole northern sky over the course of each day. Due to its location in Penticton, BC at a latitude of 49.32 degrees \citep{amiri2018chime} CHIME observes some parts of the sky continuously, and observes other parts as little as 5 minutes per day \citep{amiri2018chime}. This means that some areas of the sky are monitored as little as 0.3\% of the time each day. Thus, many more FRB repeaters could exist, but have not been detected due to censoring. Recent work by \cite{james2023modelling} and \cite{yamasaki2023}, which take into account the observational selection effects of CHIME, have suggested that the true fraction of repeaters is closer to 50\%. Surprisingly, we have reached a similar conclusion through an entirely independent approach that we present here.

In this paper, we propose a principled and interpretable statistical model to predict whether new FRB bursts are repeaters or non-repeaters. Our method uses the morphological characteristics of FRBs (e.g., bandwidth and peak frequency) as inputs, and is based on weighted logistic regression for imbalanced data sets, the details of which are described in Section~\ref{sec:methods}. In short, our classification algorithm can be used to identify potential FRB repeater candidates. Specifically, given a new FRB source for which only one burst has been recorded, the algorithm will provide the probability that this FRB source is actually a repeater. Having this kind of predictive tool could be useful for follow-up observations. For example, if a FRB source is given a high probability of being a repeater, but is in an area of the sky observed only 5 minutes per day by CHIME, then one could allocate time at other telescopes to observe it more regularly, or one could search through archival data from other telescopes to find previous bursts. While developing our prediction method and algorithm, using techniques entirely independent  from \cite{james2023modelling} and \cite{yamasaki2023}, we arrive at a similar, albeit very tentative, conclusion about the true percentage of repeating FRBs.

Creating classification and prediction algorithms for FRBs is notably difficult because of the unique structure of the data. There are at least two major challenges to overcome:
\begin{itemize}
    \item \textit{Challenge 1:} The observed number of repeaters is significantly outweighed by the observed number of non-repeaters. This imbalance in the data leads to biased inference when it is not taken into account. Some studies do not account for the imbalance \citep[e.g.,][]{chen2021}, while others have addressed the imbalance through resampling techniques \citep{yang2023}. However, the resampling approach does not account for the intrinsic reduction of  variability of features in the repeater subpopulation. 

    \item \textit{Challenge 2:} For training data, the labels for FRB repeaters are almost entirely certain, but the labels for FRB non-repeaters are not. Any non-repeater FRB may actually be a repeater that we have not yet seen repeat. This corresponds to a mislabelling problem in the training data for any classification algorithm --- i.e., there may be repeaters that are wrongly labeled as non-repeaters. Statistical approaches to mitigate this usually rely on modeling the probability of an error in labeling \citep{nagelkerke2015estimating,
hung2018robust} which, in the FRB classification case, is impossible.

\end{itemize}

Previous methods to predict and/or classify repeating FRBs have used black-box machine-learning approaches \citep[e.g.,][among others]{chen2021, luo2022, yang2023, Zhu-Ge2023}. 
However, these previous methods  inadequately handle some of the challenges posed by the particular characteristics of FRB data. For instance, \cite{chen2021}, \cite{luo2022} and \cite{luob2022} consider sub-bursts and repeater bursts as independent data points  and they do not differentiate between them when creating training and test data. Assuming independence makes it possible to split bursts from the same source and put them into the training and the test data. However, this approach will artificially i) increase the similarity between the training and test sets, since FRBs from the same source will have some dependency and, consequently, ii) enhance the model's classification performance because it is easier to identify a repeater source in the test data once one or more of its bursts have been used in  the training data. Consequently, the  use of sub-bursts from the same source in test and training data exaggerates the accuracy of the model's predictions.

The method that we propose accounts for the imbalance between the number of repeaters and non-repeaters by weighting differently the information contained in each observation. This approach relies on a tuning parameter that represents the true proportion of repeater FRBs in the universe. While a precise value is elusive, our analysis suggests that the model is robust to values of this tuning parameter between 50\% and 60\%.
Our model is also able to identify which of the non-repeaters to-date are most likely to repeat. This information can be used for strategic and efficient monitoring of the sky.

Our paper is organized as follows. First, we introduce our data selection procedure, including which FRBs features are used in our analysis (Section~\ref{sec:data}). Next, we present the proposed method of weighted logistic regression for imbalanced data sets, describe training, validation, and test data sets, and introduce the tuning parameter $\tau$ (Section~\ref{sec:methods}). We then present our results (Section 4), and conclusions and directions for future work (Section 5).

\section{Data}\label{sec:data}

We construct a statistical prediction model that will identify repeaters  based on their morphological features. Thus, we need a training set and a validation set with known labels (i.e., which FRBs are non-repeaters and which are repeaters, albeit with the caveats described in the previous section) to determine the accuracy of our methodology. Once our statistical model is trained and validated, then we can apply it to a separate test set of data. 

\subsection{CHIME/FRB Catalog}

The majority of our FRB data comes  from the first catalog of the Canadian Hydrogen Intensity Mapping Experiment \citep[CHIME;][]{chime_2021,err_chime_2023}, hereafter referred to as Catalog 1.
CHIME is a transit radio telescope operating across the 400--800MHz range and located on the grounds of the Dominion Radio Astrophysical Observatory in British Columbia, Canada. The original intention for CHIME was to map neutral hydrogen gas as a measure of dark energy, but the large collecting area, wide radio bandwidth, and powerful correlator make it an excellent instrument for the detection of FRBs as well.

The CHIME/FRB Collaboration's software pipeline searches for FRBs in real time \citep{amiri2018chime,amiri2022overview}. The first FRB catalog recorded 536 fast radio bursts between  2018 July 25,  and 2019 July 1, and at the time increased the sample of FRB sources by more than a factor of 5~\citep{petrof2022}.
By using data from only one telescope, we lessen the potential for different observational biases. An FRB, for our purposes, is constrained to the start and end of a single FRB observation, as defined by the CHIME pipeline. CHIME/FRB's first catalog contains 474 apparent non-repeating sources and 62 bursts from 18 repeating sources. For every FRB, the catalog contains 12 parameters  that can be considered possible classification features.

\subsection{Data selection and handling}\label{subsec:data}

Our exploratory variable selection analysis confirmed that  we need only consider features that were identified as useful  in previous work~\citep[see, e.g.,][]{Pleunis_2021}, as some variables seemed to have no influence in the classification results. 
The  morphological features we consider are taken from the CHIME/FRB catalog and consist of:
\begin{itemize}
    \item the boxcar width (a measure  of the FRB burst duration  in seconds),
    \item  the peak frequency (in MHz),
    \item the intrinsic width of the FRB associated with the event in seconds,  as modeled with the fitburst pipeline~\citep[i.e., without dispersion smearing and scatter-broadening,][]{Fonseca2024ApJS}, \item  the number of sub-bursts, and
    \item the emitted bandwidth in Mhz.
\end{itemize}
 
The emitted bandwidth is obtained by taking the difference between the high and low frequencies for detection at the full width at tenth maximum, which is the width of the FRB signal at the level corresponding to the difference between frequencies at which the signal reaches 10\% of its peak intensity. Previous work has only looked at emitting bandwidth and not the peak frequency, but we consider the latter as a distinct feature. Detailed descriptions of the complete set of parameters are presented by \citet{chime_2021}. In the case of FRBs with sub-bursts, we define the corresponding FRB-specific features as the mean value across sub-bursts for each feature. 

\section{Methods}\label{sec:methods}

The proposed model is based on logistic regression ~\citep{mccullagh1989generalized,Chao2002}, which is reviewed in Section~\ref{sec:log}. In Section~\ref{sec:imbdata}, we introduce from the statistics literature methods developed to account for imbalanced data sets in logistic regression, which not only helps us address the low proportion of repeaters in the FRB sample, but also allows us to introduce a tuning parameter for the fraction of repeaters in the entire population.

\subsection{Classification by Logistic Regression} \label{sec:log}

The logistic regression is a widely used statistical model that  captures the relationship between a binary dependent variable and a set of independent features, or covariates. 
Logistic regression falls under the broader class of generalized linear models \citep[see for example, ][]{McCulloch1997} for which we assume that the information in the covariates about the dependent variable is conveyed through a linear combination of features,
\begin{equation}
    \label{eqn:linmod}
    \boldsymbol{\eta} = \mathbf{X}  \boldsymbol\beta,
\end{equation}
where $\boldsymbol{\eta}$ is called the linear predictor, $\boldsymbol{\beta}$ is a vector of model parameters,  and $\mathbf{X}$ is the covariate or design matrix. That is, each row $i$ in $\mathbf{X}$ is an FRB observation, with the first column representing the intercept in the regression model and the subsequent columns representing the covariates or features. The number of observations, or rows, in $\mathbf{X}$ is $n$, and the number of columns is $m+1$. %in which the $i$-th row contains the values of the independent variables for the $i$-th  item in the sample. 

Let $\{(y_i,\mathbf{x}_i):\; 1\le i \le n\}$ denote the sample of size $n$ containing the observed FRBs in which
\begin{equation}
    y_i = \begin{cases}
      1 & \text{if the sample $i$ is from a repeater FRB source}\\
      0 & \text{if the sample $i$ is from an apparently non-repeating FRB source} \\
    \end{cases}
\end{equation}
and $\mathbf{x}_i \in \mathbf{R}^{(m+1) \times 1}$ is the column vector of observation $i$ that contains $m$ features. While the description is valid for all $m<n$, for the FRB model we selected five features ($m=5$): the boxcar width, peak frequency, intrinsic width from fitburst, number of sub-bursts, and emitted bandwidth, as described in Section~\ref{subsec:data}.

The logistic regression model assumes that each $y_i$ follows a Bernoulli distribution
\begin{equation}
    y_i \sim \text{Bern}(\pi_i(\boldsymbol\beta))
\end{equation}
where 
\begin{equation}
     \pi_i(\boldsymbol\beta) =  \frac{e^{\eta_i}}{1+e^{\eta_i}}
\label{pii}
\end{equation}
and \begin{equation}
    \label{eta}
\eta_i=\mathbf{x}_i^T \boldsymbol\beta=\beta_0 + x_{1i}\beta_1 + x_{2i}\beta_2 + x_{3i}\beta_3 + x_{4i}\beta_4 + x_{5i}\beta_5,\end{equation}
where  $x_{ji}$ is the $j$-th feature or covariate of observation $i$.

The inference for $\boldsymbol{\beta}$ is based on the likelihood function
\begin{equation}
    \mathcal{L}(\boldsymbol\beta|\text{Data}) =   \prod_{i=1}^n  \left [\pi_i(\boldsymbol\beta)^{y_i} (1-\pi_i(\boldsymbol\beta))^{(1-y_i)} \right ]
\end{equation}
where $\text{Data}=\{(y_i,\mathbf{x}_i): \; 1\le i\le n\}$. Estimates of $\boldsymbol{\beta}$ can be obtained through the maximum likelihood estimator, 
$\hat{\boldsymbol\beta} = \arg\max_{\boldsymbol\beta} \mathcal{L}(\boldsymbol\beta|\text{training data})$, and cross-validation can also be used to further test and validate the classifier model. 

With $\hat{\boldsymbol{\beta}}$ in hand, the logistic regression model (or classifier) can then be used to classify any new FRB with observed features are $\mathbf{x}^*=(x_1^*,\ldots,x_5^*)$ as a repeater or non-repeater. That is, one uses $\hat{\boldsymbol\beta}$ and $\mathbf{x}^*$ to obtain $\eta_i^*$ and $\pi_i^*(\hat{\boldsymbol\beta})$ using equations \eqref{eta} and \eqref{pii}. $\pi_i^*(\hat{\boldsymbol\beta})$ is the probability $P(\text{$i$-th FRB is a repeater})$
%, which is sometimes called the positive class, and for a binary classification, $P(\text{repeater}) + P(\text{non-repeater}) = 1$. 
Although the model returns a probability value, in practice a threshold is used to enable metrics for the classification performance. The most common threshold is $0.5$, and we use the  criteria $\pi_i^*(\hat{\boldsymbol\beta}) >0.5 $ to predict that the $i$-th FRB is a repeater.

One drawback of logistic classifiers is that they do not perform well with imbalanced data sets, i.e. situations in which the number of cases ($Y_i=1$) is vastly different from the number of controls ($Y_i=0$) \citep{imbalance2009, Luque2019, empiricalsamp2022}. In the next section, we describe potential solutions to alleviate this challenge.%s posed by  particular characteristics of FRB data that prevent us from using a standard logistic regression. 

\subsection{Imbalanced Data} \label{sec:imbdata}

In the statistics literature, imbalanced data refers to data for which --- due to  observational bias, poor sampling design, or population structure --- the number of samples from different subpopulations are substantially different. The imbalance may be extrinsic, i.e.,  an unknown mechanism unrelated to the model describing the samples hides the real subpopulation ratio \citep{empiricalsamp2022}. The imbalance may instead (or also) be due to exogenous sampling, i.e., the imbalance happens during the data sampling process \citep{manski1977}. The imbalance may also be intrinsic in situations where only a small fraction of items belong to a subpopulation.%, e.g. in credit card fraud. 

In this study, we have an imbalanced data set between repeaters and non-repeaters~\citep{chime2023}. The number of bursts from non-repeaters vastly dominates the number of bursts from repeaters. Whether this imbalance is extrinsic, intrinsic, or due to exogenous sampling is still up for debate. Intuitively, it could be due to the fact that some of the putative non-repeaters  in the sample would yield another burst if they were monitored for a longer period of time. The idea that the actual percentage of repeaters in the population is much larger than the one computed directly from the samples collected to date is gaining momentum. For instance, \cite{james2023modelling} recently proposed a power-law model for the FRB population as a function of redshift, showing that at least 50\% of bursts in CHIME's first catalog should come from intrinsic repeaters. Independently, \cite{yamasaki2023} suggests the necessity of correcting the observed source count according to declination to adequately model the evolution of repeaters and non-repeaters, in particular the transition of apparent non-repeaters when they are observed to repeat. A model population incorporating this correction results in an inferred repeater fraction that exceeds  50\%.

Regardless of the cause of imbalanced data for FRBs, it is this imbalance that hinders the direct application of a logistic regression model to classify FRBs. Statistically, the data imbalance leads to biased estimates of the parameters and therefore biased predictive probabilities~\citep{mccullagh1989generalized, king2001, goorbergh2022, imbalance2009, maalouf2011}.

There are two general approaches to alleviate the imbalance \citep{imbalance2009}. The first is to make the data set more balanced by using a subsample of the larger population. In our case, this would mean using all the repeaters but only a subsample of non-repeaters, and then proceeding as if that were the data in hand ~\citep[see][for an empirical study of these methods]{empiricalsamp2022}. This approach only makes sense, however, if one knows the true proportion of repeaters in the population. The second approach to make the data set more balanced is to over-sample the smaller class. However, such an approach introduces additional variability in the analysis from the subsampling procedure \citep[see][for a discussion]{goorbergh2022}.

We therefore favor a third approach that applies weights to re-balance the data. This is done using the method of {weighted logistic regression} for imbalanced data sets developed by \cite{king2001}, \cite{kingzeng2001}, \cite{King2003}, \cite{maalouf2011} and \cite{maalouf2014}. The weights are defined by introducing a parameter $\tau$ to denote the proportion of repeaters among the whole population of FRB sources.  We assume $\tau$  is different than the observed fraction in the sample. %, and use the method of {weighted logistic regression} for imbalanced data sets developed by \cite{king2001,kingzeng2001}, \cite{King2003}, \cite{maalouf2011}, and \cite{maalouf2014} to weight the data points corresponding to repeater and non-repeater sources. 
For simplicity, we only consider the case of $\tau$ as a fixed fraction, independent of redshift, luminosity, etc.

In weighted logistic regression,  a weight $w_i$ is assigned to the $i$-th sample 
 % weighted logistic regression
~\citep{king2001,kingzeng2001, King2003, maalouf2011, maalouf2014}. Following \cite{kingzeng2001},  the weight for the $i$-th sample is 
\begin{equation}
    w_i  =\delta_ry_i + \delta_{nr}(1-y_i)
\end{equation}
where \begin{equation}
    \delta_{nr} =  \frac{1 - \tau}{1 - \Bar{y}},\qquad \delta_r = \frac{\tau}{\Bar{y}},
\end{equation}
where $r$ and $nr$ stand for repeater and non-repeater respectively, and $\Bar{y}$  is the number  of repeaters in the data divided by the total number of observed sources. When $y=1$ (a repeating FRB), $w=\delta_r=\frac{\tau}{\bar{y}}$. In this way, $\delta_r$ could be interpreted as the correction factor for the proportion of repeaters in the sample. Note that each weight $w_i$ depends on the ratio $\tau$. The latter is not known exactly, and so we first treat $\tau$ as a tuning parameter, trying various values of $\tau$ between 0 and 1. Ultimately, though, we rely on a combination of the evidence provided by \cite{yamasaki2023}, who state that $\tau \in (0.5,1]$, and the results of our initial exploration of $\tau$ settle on the value  $\tau=0.55$. This will be discussed further in Section~\ref{sec:resultstau}.

The resulting (weighted) logistic regression log-likelihood is then
\begin{equation} \label{eq:weightlike}
    \log\mathcal{L}(\boldsymbol\beta|\text{Data})  = - \sum^N_{i=0}w_i\log(1+e^{(1-2y_i)\eta_i}),
\end{equation}
where $\eta_i$ is defined as in Equation \eqref{eta}.

The maximum likelihood estimator for $\boldsymbol{\beta}$ derived from Equation \eqref{eq:weightlike} incurs an increase in bias and variance, given the number of samples, as discussed by \cite{kingzeng2001} and \cite{maalouf2014}. The introduction of weights causes the estimators to become biased as proved by~\cite{mccullagh1989generalized} and in appendix A of~\cite{kingzeng2001}. However, one can correct for this; we use the maximization algorithm that corrects for bias and reduces variance using the methods proposed by \cite{maalouf2014}. Using their approach, we obtain estimates of $\boldsymbol{\beta}$ from our training data, and use cross-validation to create our classifier.

We note that the mislabeling of a repeater as a non-repeater simply because it was not observed for long enough can produce serious biases in any supervised learning-type analysis. This is a caveat that has not been addressed in previous work  ~\citep{luo2022, luob2022, yang2023}. The introduction of the true proportion $\tau$ allows us to account for the uncertainty associated with the potential mislabeling of repeating sources as non-repeaters.

\subsection{Training and Test Data} \label{sec:train}

Ideally, we would use multiple data sets to study the statistical performance of the proposed model. However, this is not possible because we have only one data set at our disposal. Thus, we instead use cross-validation; we create multiple data sets by repeatedly splitting the available data set into a training set and a cross-validation set. For each split, we fit the model on the training data and then classify the cross-validation data as repeaters or non-repeaters based on the fitted model. The classification is done by evaluating the the probability $\pi_i$ in Eq.~\eqref{pii}, which returns the probability of belonging to one of the two possible classes. 

To assess the performance of a classifier, it is customary to use a classification threshold of 50\%. More precisely, if the estimated probability of an FRB being a repeater is greater than 50\%, i.e., $Pr(Y=1)\geq 0.5$,  then we classify the FRB as a repeater. Changing the threshold has an effect on the performance metrics, but the probability itself is not affected. 

Instead of a single classifier, our approach yields an ensemble of classifiers, since every data split will yield a different set of parameter estimates. The procedure of data-splitting and model-fitting is repeated 500 times, which provides uncertainty quantification for the model's classification accuracy. While more replicates are possible, the results remain similar due to the small number of repeaters in the sample.  When performing the data-splitting, we carefully avoid situations in which repeated bursts from the same source appear in both the training and test data sets. 
In other words, repetitions from the same FRB source will all be included in the training set or will all be included in the cross-validation set at each data split.

\begin{figure}
    \centering
    \includegraphics[height=0.4\textwidth]{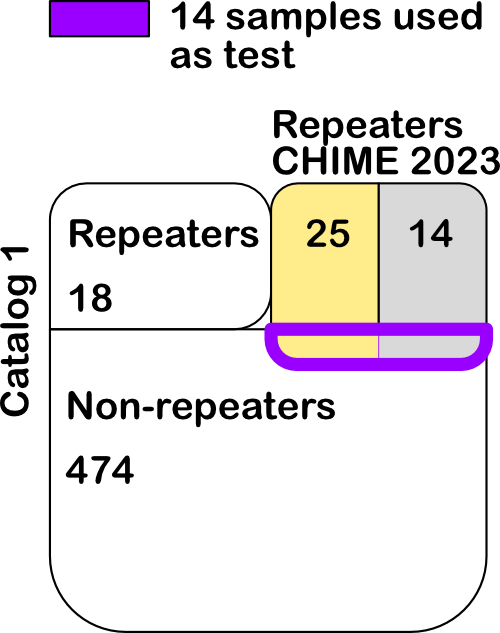}
    \caption{Venn diagram describing our training/validation set and ``test'' set. The training/validation set is comprised of the white, yellow, and grey regions, minus the region delimited by the purple outline. The latter is the ``test'' set of 14 FRBs and  the performance of our model on these is shown in Fig.~\ref{fig:test}.}
    \label{fig:venn}
\end{figure}

The number of confirmed repeater sources in Catalog 1 is only 18, which is much smaller than the  474 non-repeaters (together shown as white regions in Fig.~\ref{fig:venn}) --- this limits the number of repeater sources in the test data. Since the release of Catalog 1, the CHIME/FRB collaboration has published an additional 25 newly discovered repeaters (the ``Gold sample'', shown as the yellow region in Fig.~\ref{fig:venn} and in ~\citealt{chime2023}). For the purpose of our interpretation, we refer to these as ``real repeaters''.

\cite{chime2023} also include 14 newly discovered repeater candidates (the ``Silver sample'', shown as the grey region in Fig.~\ref{fig:venn}), which have a lower confidence of being repeaters than the Gold Sample, but have a higher confidence than the non-repeaters. We include the Silver sample as repeaters in our training and cross-validation splits.

Out of these 39 sources in the Gold and Silver samples, we exclude a subset of 14, which cross-match in Catalog 1 and the Repeaters catalog, from training and cross-validation. These 14 FRB sources are used as our test set (delimited by the purple outline in Fig.~\ref{fig:venn}). The remaining 25 repeaters and non-repeaters are added to the training/cross-validation set.

\subsection{Selection of $\tau$} \label{sec:tau}

The statistical model that we consider for classification depends on the parameter $\tau$, which is not directly estimable from the data, or from theory.

In this paper, we treat $\tau$ as an unknown independent quantity and  create a sequence of its possible values. For each value of $\tau$, we train the proposed logistic regression classifier by splitting randomly 500 times the training and cross-validation sets, as  described above. Together, the individual classifiers trained by the random splits form an \textbf{ensemble of classifiers}. For every value of $\tau$, a different maximum likelihood estimator is obtained and a different classification for a newly observed FRB source is generated.

As mentioned previously, the logistic regression model gives a probability $\pi_i$ of belonging to one of the classes (either repeater or non-repeater), but this is not useful when measuring the performance of the model, as our response, $y_i$, is binary in nature. So it is necessary to apply a decision threshold --- we use a standard value of 50\%. The choice of this threshold is independent of $\tau$ or the imbalance of the data, and it only serves as a decision boundary to obtain a binary response. The standard value assumes that there is no preference for repeaters or non-repeaters, and allows for better analysis of the performance of the logistic regression model.

When summarizing the performance of the model for each $\tau$, we use the traditional metrics of accuracy, precision, recall, and $F_{1}$-Score. These metrics are derived using empirical frequencies computed from the proposed model's performance on multiple splits of the training and cross-validation data. These metrics are defined as follows:  
\begin{itemize}
    \item \textit{Accuracy:} the proportion of correct predictions among all predictions, i.e., the proportion of correctly classified FRBs as repeaters and non-repeaters, among all classifications:
    \begin{equation}
        \text{Accuracy }=\frac{TP+TN}{TP+TN+FP+FN}
        \label{eq:accuracy}
    \end{equation}
    where $TP$, $TN$, $FP$, and $FN$ are, respectively, the number of true positives, true negatives, false positives and false negatives. 
    \item \textit{Precision:} proportion of  positive predictions that are actually correct, i.e., the proportion of FRBs classified as repeaters that are real repeaters:
    \begin{equation}
        \text{Precision }=\frac{TP}{TP + FP}
        \label{eq:precision}
    \end{equation}
    
    \item \textit{Recall:} proportion of true positives that have been correctly 
    predicted, e.i., the proportion of real repeaters that are correctly identified as repeaters:
    \begin{equation}
        \text{Recall }=\frac{TP}{TP + FN}
        \label{eq:recall}
    \end{equation}
    
    \item \textit{$F_{1}$-Score:} a measure that combines Precision and Recall. The $F_{1}$-Score is the harmonic mean of the two, and can be interpreted as an optimal number of true positives without introducing too many false negatives or false positives. It can also be written as:
    \begin{equation}
        F_{1}\text{-Score} = 2\frac{\text{Precision}\times\text{Recall}}{\text{Precision} + \text{Recall}}
        \label{eq:F1}
    \end{equation}
    The $F_{1}$-Score is particularly useful in cases where the data set is imbalanced.
\end{itemize}

After training an ensemble of classifiers\footnote{Example implementation is available at \url{https://github.com/alfa33333/RE-FRB}} for every $\tau$, we analyse the above performance metrics (Sec.~\ref{subsec:softlim}) and combine this information with literature estimates to settle on a single value of $\tau$.

%%%%%%%%%%%%%
\section{Results} \label{sec:FRBclass}

\subsection{Example demonstrating weighted logistic regression for imbalanced data sets} \label{sec:resultstau}

\begin{figure}
    \centering
    \includegraphics[width=0.49\textwidth]{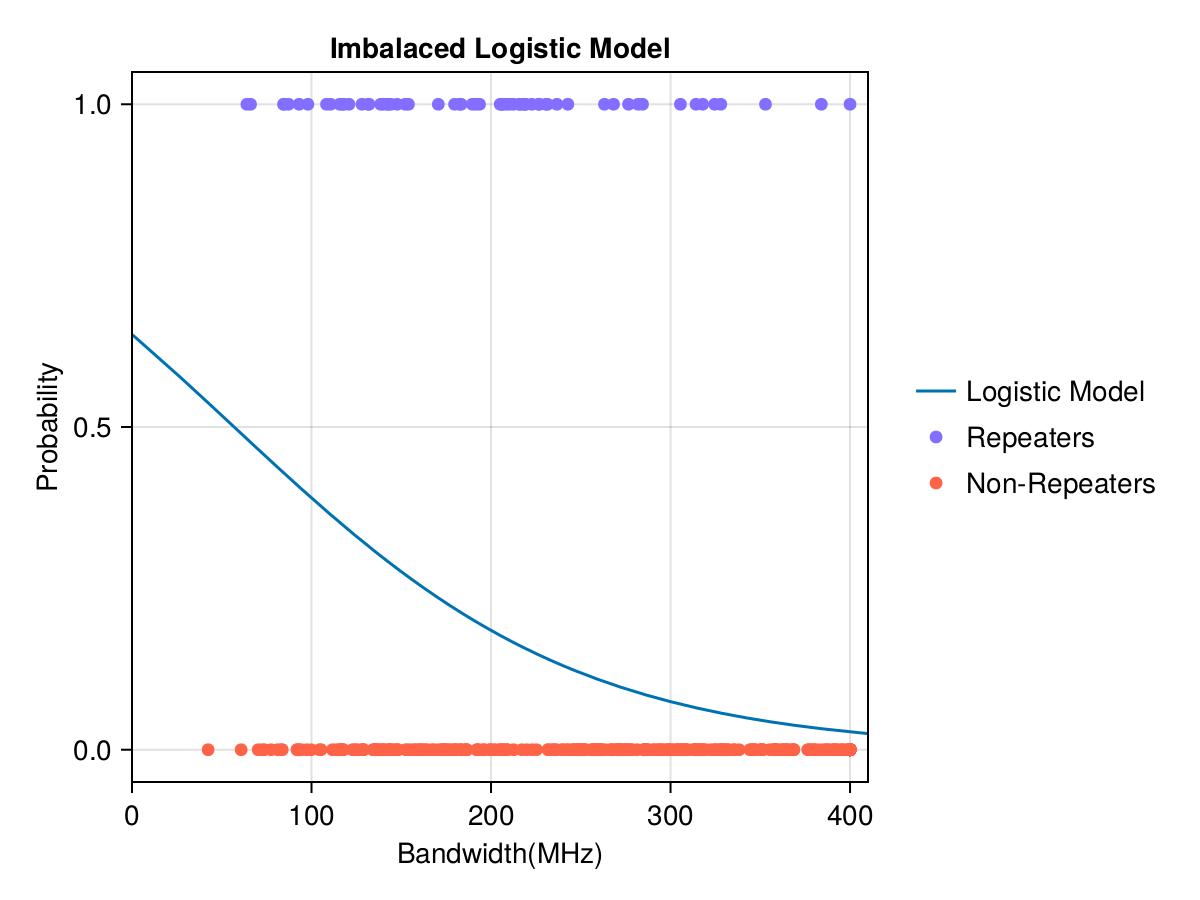}
    \includegraphics[width=0.49\textwidth]{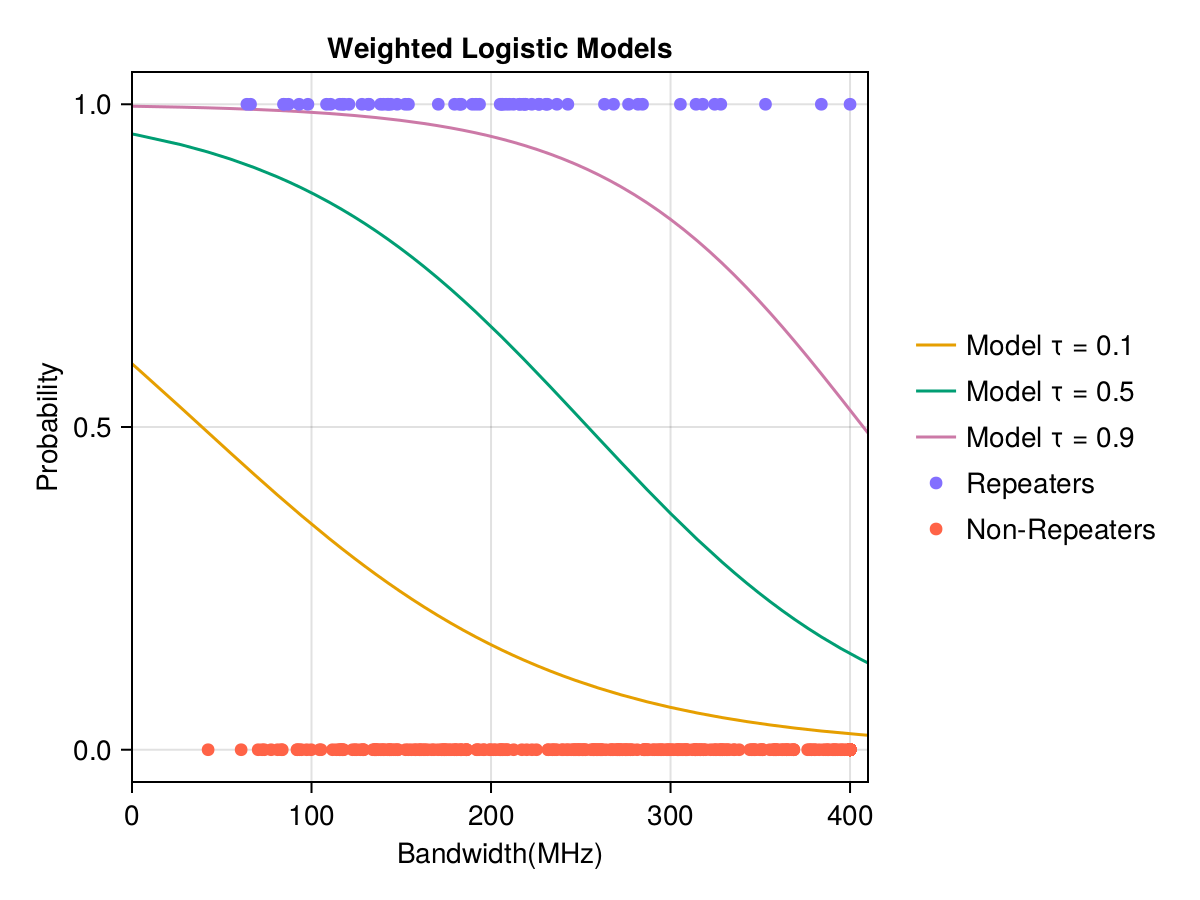}
    \caption{Example of a logistic classification model using the bandwidth of each FRB as a feature and the complete catalog as training data. The purple and orange dots represent the catalog 1 samples, and the $y$-axis is the probability of belonging to one of the binary classes, where 1.0 is a 100\% or the repeater class, and 0.0 is 0\% or the non-repeater class. As the data have a binary response, the samples are assigned to 0\% or 100\% depending of the assigned label of repeater or non-repeater. \textit{Left:} The classical logistic model ignores the imbalance in the data, and produces the fit shown as a solid blue line. Based on this model fit, and assuming a threshold probability of 50\% to classify any future data as repeaters, only a few samples would be correctly classified. \textit{Right:} The weighted logistic regression uses the rare events approach and an assumed value of $\tau$ (Section~\ref{sec:imbdata}), resulting in the fit shown as solid lines ($\tau=0.1$ in orange, $\tau=0.5$ in green, and $\tau=0.9$ in pink). 
    }
    \label{fig:cat1}
\end{figure}

In Figure~\ref{fig:cat1}, we compare  classifications  produced by traditional logistic regression that does not account for imbalance in the data (left) to that produced by weighted logistic regression (right). In this example, and only for ease of illustration, we use the bandwidth as a predictor and Catalog 1 as the training data. In the right-hand side of Figure~\ref{fig:cat1}, we also show how the logistic curve changes as  $\tau$ takes the values  $0.1$, $0.5$ and $0.9$. For $\tau=0.1$ the fitted logistic curve (orange line) is very similar to the fit obtained  without any correction. When $\tau=0.5$ the logistic fit changes shape, and predicts higher probability $\mathrm{Pr}(Y=1)$ (i.e., predicts a repeater) for FRBs with lower bandwidths. When $\tau=0.9$ we can see that most of the FRBs will be predicted to be repeaters, which is not surprising since the information  provided by $\tau$ is that 90\% of all FRB sources in the Universe are repeaters. Clearly, the choice of $\tau$ has a strong effect on the model. When judging the accuracy of the model, one must consider the rate of mislabeling a non-repeater, which relates with $1-$Precision, or mislabeling a repeater, which relates to $1-$Recall. In general, we expect both type of errors to be non-zero and impossible to simultaneously optimize \citep[see][and references therein]{craiu2008choosing}. One must therefore consider the effect of choosing a value of $\tau$ on both types of errors simultaneously.

\subsection{Ensemble of Classifiers: finding a soft-limit for $\tau$} \label{subsec:softlim}

\begin{figure*}[t!]
    \centering
    \includegraphics[width=0.49\textwidth]{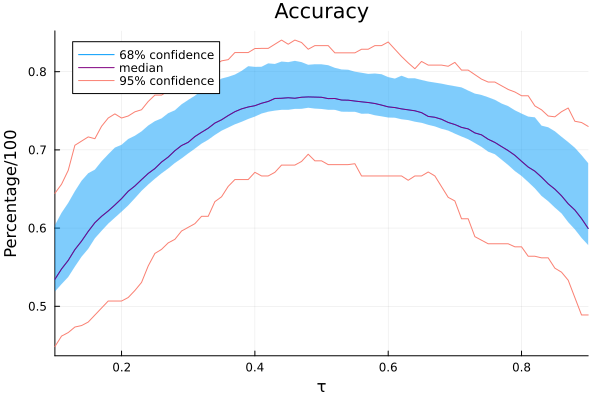}
    \includegraphics[width=0.49\textwidth]{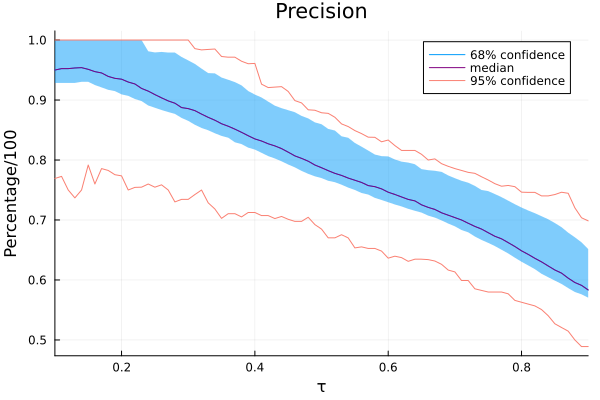}
    \includegraphics[width=0.49\textwidth]{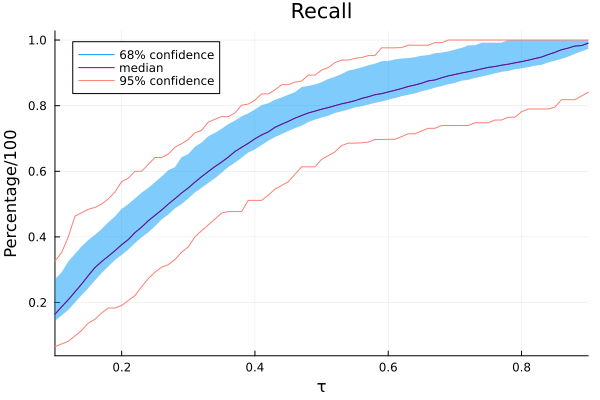}
    \includegraphics[width=0.49\textwidth]{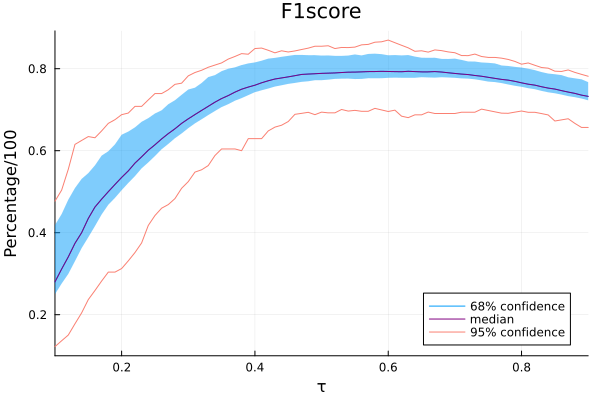}
    \caption{Performance metrics for the ensemble of classifiers, as a function of $\tau$. Values on the vertical axis represent the proportion for each metric (i.e., equations~\ref{eq:accuracy}-\ref{eq:F1}). Accuracy (the proportion of correctly classified FRBs as repeaters and non-repeaters, among all classifications) is shown in the upper left, and is maximized for $0.4\leq\tau\leq0.6$. Precision (the proportion of FRBs correctly classified as repeaters out of all repeater predictions) is shown in the upper right, and decreases with increasing $\tau$. Recall (the proportion correctly identified repeaters out of all real repeaters) is shown in the lower left, and increases with increasing $\tau$. $F_{1}$-Score (the average rate between Precision and Recall) is shown in the lower right, and appears to be maximized and stable in the approximate range $0.4\leq\tau\leq0.6$, similar to the accuracy. Precision and Recall appear to be inversely related. }
    \label{fig:repeaters}
\end{figure*}

In Figure~\ref{fig:repeaters}, we show the performance metrics and their estimated confidence intervals, as a function of $\tau$, listed in Section~\ref{sec:methods} for our ensemble of classifiers. The bold purple line represents the median, the blue-shaded regions represent the 68\% confidence intervals, and the orange lines are the 95\% confidence intervals. 
The most important feature of Figure~\ref{fig:repeaters} is that the range $0.4\leq\tau\leq0.6$ gives the most reliable model.

Recall from equation~\ref{eq:F1} that the $F_{1}$-Score provides an average of Precision and Recall and thus can be considered an overall measure of performance. Similarly, Accuracy (equation~\ref{eq:accuracy}) measures how often the model correctly predicts the output, regardless of the true value. As illustrated in the upper left and lower right panels of Fig.~\ref{fig:repeaters}, both Accuracy and $F_1$-Score show an optimal range of approximately $0.4\leq\tau\leq0.6$. 

For values lower than $\tau=0.4$, the Accuracy, Recall, and $F_{1}$-Score return a lower percentage of correctly classified samples. Although the Precision (upper right panel, Figure~\ref{fig:repeaters}) is high for low $\tau$, this is merely a consequence of the majority of the data being non-repeaters; the Precision (Equation~\ref{eq:precision}) only considers the performance as measured on positives (i.e., repeaters).  Since the majority of the data consists of negatives (i.e., non-repeaters), the method is more likely to misclassify a negative. We also note that Precision is insensitive to the number of repeaters correctly classified. For example, we could achieve 100\% Precision by having just one repeater correctly classified and zero misclassified non-repeaters.  

Recall (lower left panel of Figure~\ref{fig:repeaters}) shows the ability of our model to correctly identify repeaters. Ideally we want to achieve  100\% Recall; based on Equation~\eqref{eq:recall}, achieving 100\%  means that every repeater is correctly identified as a repeater. Our model indicates that a high value of $\tau$ is needed for such a result. Such a high value of $\tau$ would imply a belief  that almost all FRBs in the universe are repeaters. %and depends on the amount of contamination present in the sample.

For values higher than $\tau=0.6$, the Accuracy and $F_1$ Score decline. Although it is possible to correctly classify all repeaters with $\tau = 0.8$, using $\tau=0.6$ still correctly classifies all repeaters in the catalog for some of the test splits. While $\tau>0.6$ is possible, we are not comfortable with the implication of having such a large proportion of repeaters, as we do not have other external evidence that justifies the decrease in Accuracy and Precision from the data.

Based on the overall performance metrics in Figure~\ref{fig:repeaters}, it appears that our ensemble of classifiers performs best in the range $0.4\leq\tau\leq0.6$. Both \cite{james2023modelling} and \cite{ yamasaki2023} propose that the fraction of repeaters should be at least 50\%, by modeling the 
dispersion measure (DM) with low repeating population or correcting the source count evolution, respectively. Combining this lower limit of $\tau$ with our range, we are left with a proportion of repeaters somewhere between 50\% and 60\%. The analysis proposed in this paper is quite robust to the choice of $\tau$ in this range. For brevity, we present the results from a single value of $\tau=0.55$ for the remaining analysis.
%(a conservative choice given the credible range for this tuning parameter). 
The results corresponding to $\tau=0.5$ and $0.6$, the lower and upper limits of the range, are very similar. 
 
\subsection{Testing our model with the gold and silver ``test'' data}

\begin{figure*}
    \centering
    \includegraphics[width=0.9\textwidth]{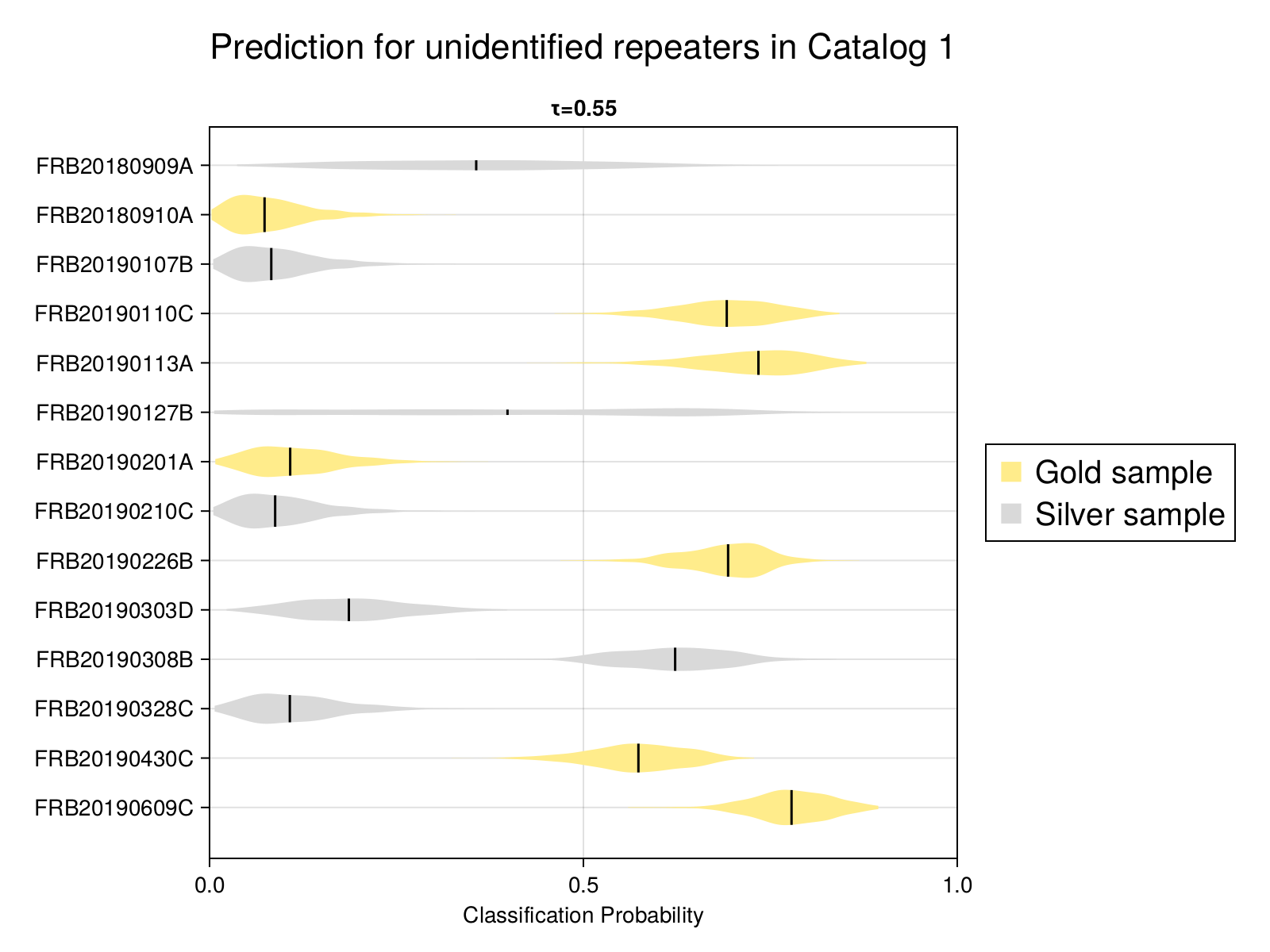}
    \caption{Violin plots of the classification probability for the ensemble of classifiers at $\tau=0.55$ for the 14 unidentified repeater sources in catalog 1. The non-repeater to repeater threshold of 50\% is shown as a gray vertical line; everything to the right of the gray line is classified as a repeater. The gold and silver colors correspond to the gold and silver samples, respectively. Each of the violin plots indicates where the majority of the distribution lies, and the dark lines inside the boxes are the medians of the distributions. The median values are enough to identify 6 out of 14 as repeaters across the different values of $\tau$. The FRB20180909A and FRB20190127B classification distributions are stretched so  thinly across the threshold that they become ambiguous to classify.}
    \label{fig:test}
\end{figure*}

We use the test data (i.e., the 14 FRB sources from the Gold and Silver samples described in Section \ref{sec:train}) to assess the performance of the ensemble of classifiers for $\tau=0.55$. The results are presented in Figure \ref{fig:test}. The y-axis shows the names provided by the Transient Name Server (TNS) for each FRB in the test set, and the x-axis shows the classification probability. If the classification probability for an FRB is higher than 0.5, then we classify the FRB as a repeater. Since each FRB is classified multiple times through the ensemble of classifiers, the results are summarized by violin plots of the predicted probabilities. The vertical black lines are the 
medians of all the prediction probabilities.  We choose to present the median instead of the mean because the distribution of predicted probabilities is not symmetric and the mean may be misleading. The distributions are colored gold and silver, to indicate which FRBs belong to the Gold and Silver Samples, respectively. 

Of the 14 samples in our test set, six (four Silver FRBs and two Gold FRBs) are unambiguously identified as non-repeaters because the entire distribution is below the threshold. Two FRBs remain ambiguous (FRB20180909A and FRB20190127B, both Silver FRBs) due to the distribution being broad and uninformative. We found that this result is consistent across three values for $\tau \in\{0.5, 0.55, 0.6\}$. Notably, five of the six samples identified as repeaters are from the Gold sample.

The remaining FRBs in Figure~\ref{fig:test} are predicted to be repeaters, with only the tail of the distributions crossing the threshold to the non-repeater region. Note that one could find several ways to make use of the distribution of classification probabilities. For instance, in the case of FRB20190609C, all the ensemble models correctly predict it to be a repeater and  the median of the classification probabilities value is significantly higher than 0.5, which strongly recommends it for future monitoring.    In the case of FRB20190127B (sixth from the top in Figure~\ref{fig:test}) the set of ensemble models that predicts it as a repeater is barely outnumbered by the complement set.  An observer who is interested in finding new repeater signals might consider the ambiguity in the evidence and choose to continue monitoring this FRB.

\begin{figure*}
    \centering
    \includegraphics[width=0.9\textwidth]{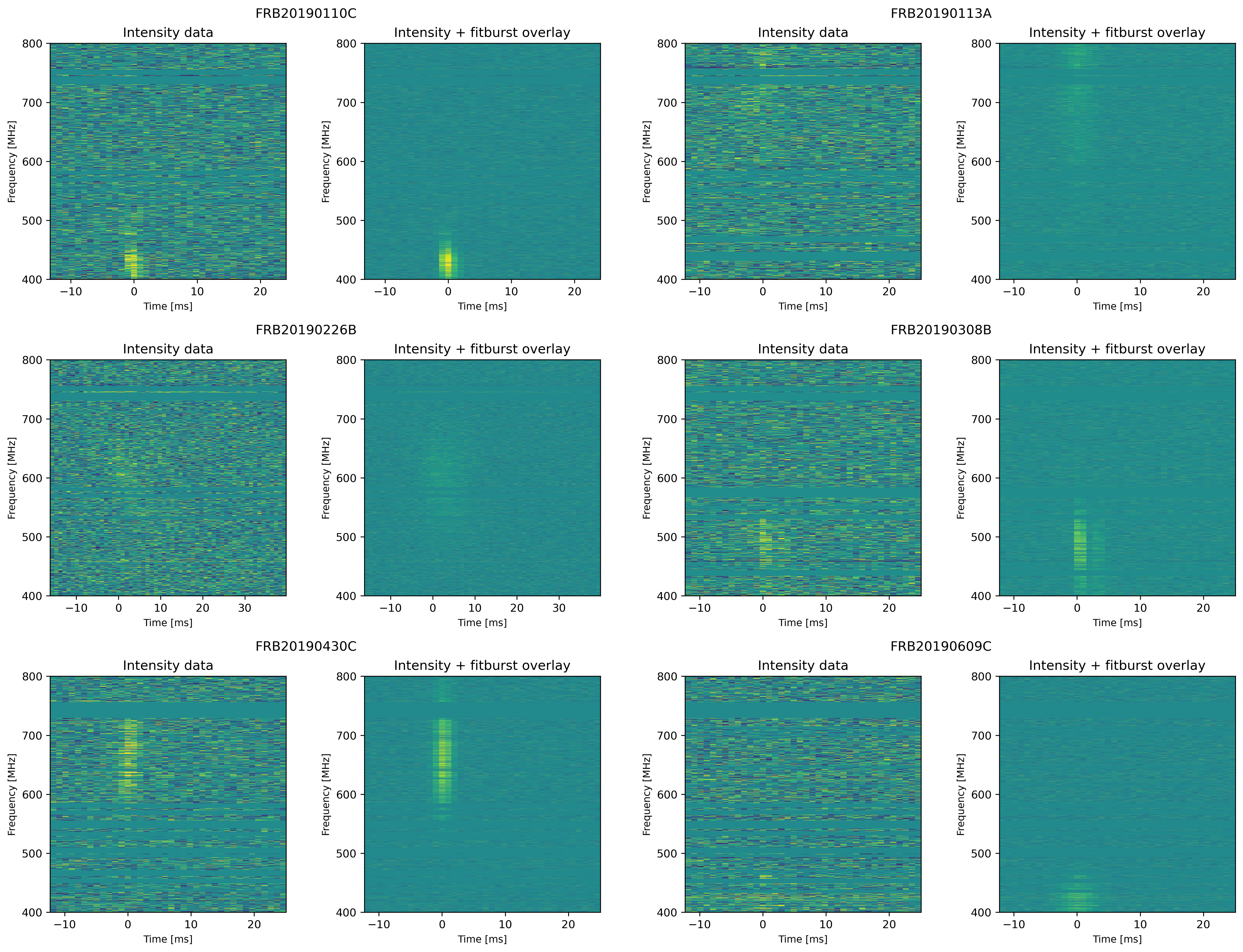}
    \caption{Dynamic spectra or waterfall plots for the six FRB burst from the test set, from~\citet{chime_2021}, which are classified as repeaters by the ensemble classifier (i.e., FRB20190110C, FRB20190113A, FRB20190226B, FRB20190308B, FRB20190430C, and FRB20190609C from Figure~\ref{fig:test}). We present two plots for each sample; at the left is the intensity data and at the right is an overlay of the fitburst model to enhance the visual shape of the FRB. These bursts show evidence of being a repeater: wider widths, downward drifting, or narrow emitting bandwidths. This is in contrast with bursts from Figures~\ref{fig:listincorrect} and ~\ref{fig:listambiguous}, which do not show repeater characteristics.}
    \label{fig:listcorrect}
\end{figure*}

\begin{figure*}
    \centering
    \includegraphics[width=0.9\textwidth]{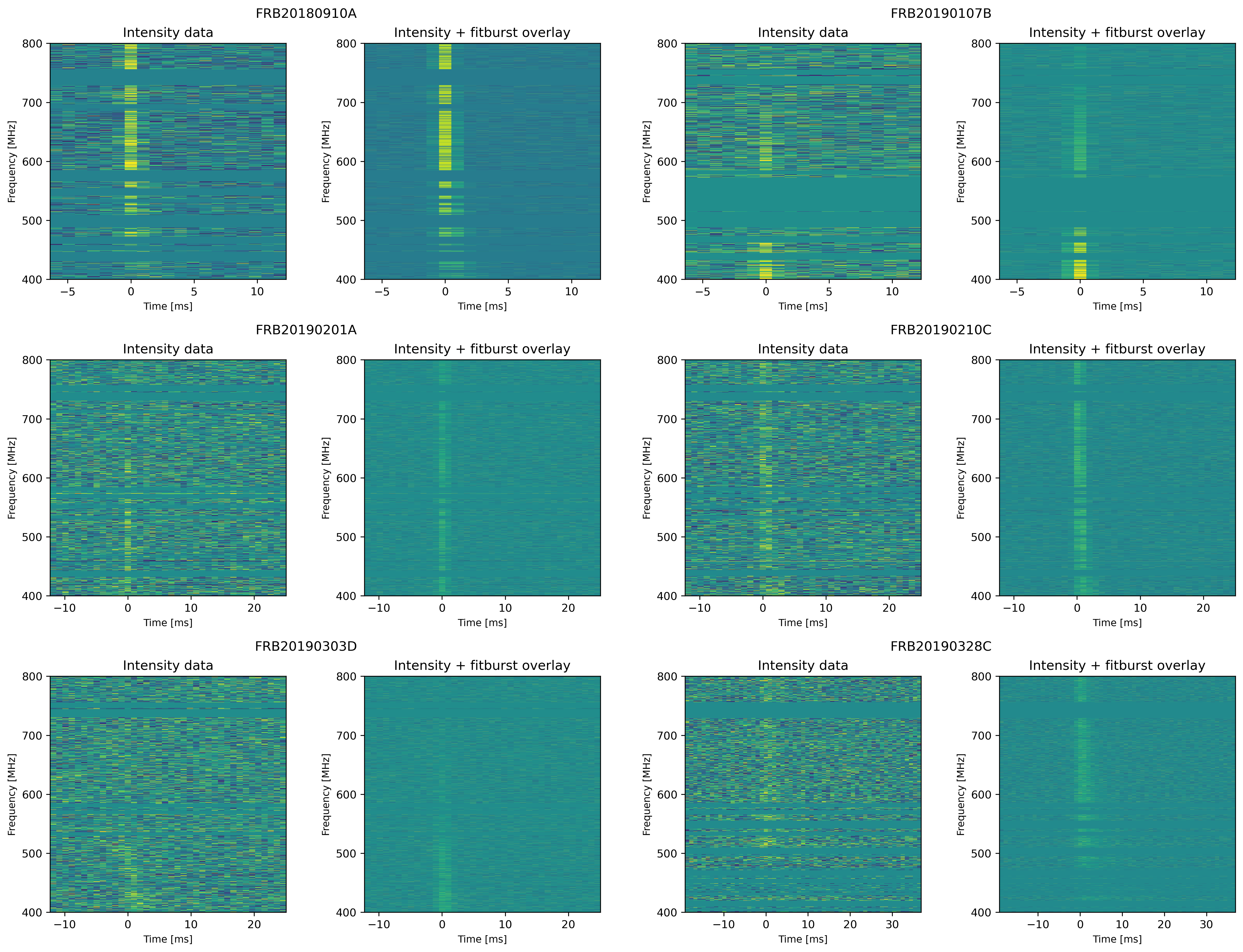}
    \caption{Dynamic spectra or waterfall plots for the six FRBs from the test classified as non-repeaters by our method (i.e., FRB20180910A, FRB20190107B, FRB20190201A, FRB20190210C, FRB20190303D, FRB20190328C  from Figure~\ref{fig:test}). These FRBs' morphologies appear  similar to non-repeaters.}
    \label{fig:listincorrect}
\end{figure*}

\begin{figure*}
    \centering
    \includegraphics[width=0.9\textwidth]{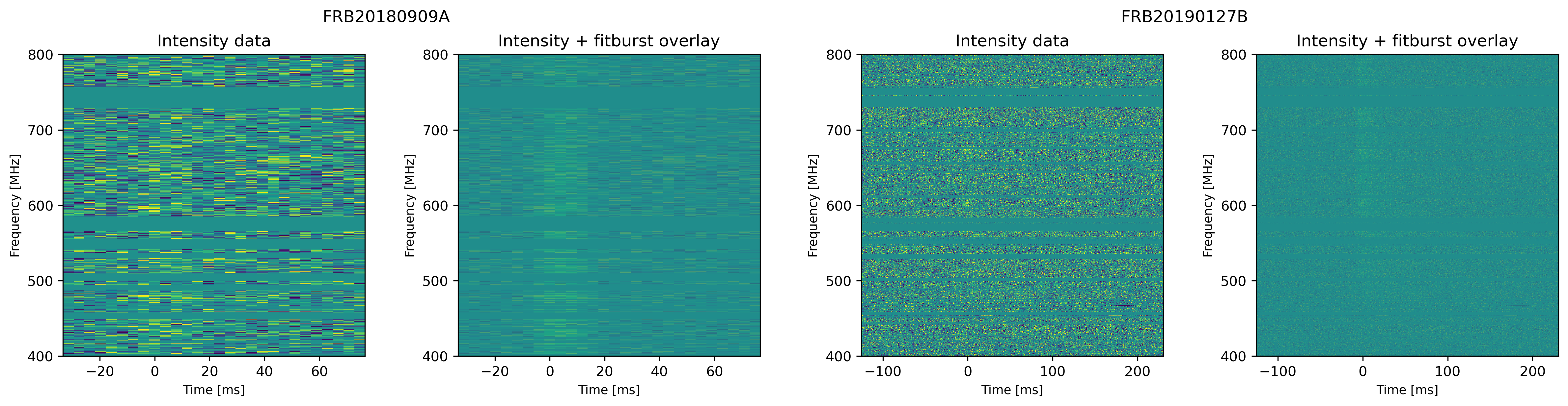}
    \caption{Dynamic spectra or waterfall plots of the two FRBs in the test set (FRB20180909A and FRB20190127B) that are  ambiguously classified using our method.}
    \label{fig:listambiguous}
\end{figure*}

% talk in more detail about each repeater here
At first glance, Figure~\ref{fig:test} shows that six out of 14 FRB sources in the test set were correctly labelled as repeaters. A quick and naive interpretation of this result is that our model does no better than 50\% chance at predicting repeaters. However, six out of 14 repeaters is merely a result of the 50\% threshold for classification. Had a different threshold been chosen, then more or fewer FRBs would be classified as repeaters. Moreover, the classification is not random between the Gold and Silver samples. Almost all of the Gold sample are identified as repeaters, and most of the Silver sample are not, which implies that information from the covariates is informing our model. Furthermore, the training data are imbalanced to non-repeaters, so if the classifiers give an FRB source in the test set a high probability of being a repeater, then that FRB must be quite different from the non-repeater set. The Precision in Figure~\ref{fig:repeaters} gives another measure of this idea --- approximately 20\% of non-repeaters are miss-classified at our selected $\tau$.

In Figures~\ref{fig:listcorrect}, \ref{fig:listincorrect}, and \ref{fig:listambiguous}, we strengthen our case that our classifier is doing better than chance by showing the dynamic spectra, also called ``waterfall plots'', for each FRB in the test set. We present two plots for each FRB, at the left the intensity data and at the right the intensity data with an overlay of the fitburst model for visibility purposes. FRBs classified as repeaters are shown in Figure~\ref{fig:listcorrect}, non-repeaters in Figure~\ref{fig:listincorrect}, and ambiguous FRBs in Figure~\ref{fig:listambiguous}. In these waterfall plots, we see the typical behaviour of FRB burst morphology for repeaters and non-repeaters; those with broad widths are classified as repeaters and those with a single short burst are classified as non-repeaters. 

Each of the repeating sources of FRBs within the Gold and Silver samples have a contamination rate of chance coincidence \citep[$R_{cc}$;][]{chime2023}. The latter is interpreted as a measure of uncertainty. A higher uncertainty is represented by a higher $R_{cc}$ value, and is linked with the probability of being observed by chance in the same region as another source. The Gold sample includes sources with $R_{cc} < 0.5$ while the silver sample includes sources within the range $0.5 \leq R_{cc} < 5$. This additional information provides a different perspective on our results. From the Gold sample, our method is able to correctly identify 5 out of 7 sources as repeaters, yielding an accuracy on confirmed repeaters higher than $70$\%, which is more in line with the  results generated via cross-validation. The two FRBs from the Gold sample that were miss-classified, FRB20180910A and FRB20190201A, can be considered atypical cases or outliers in their morphological features. While FRB20180910A has the largest contamination rate from Gold sample with $R_{cc} = 0.43$, FRB20190201A is similar to FRB20200120E, which shows unusual larger bandwidths and narrower widths~\citep{Bhardwaj_2021}.

\section{Conclusions and future work}\label{sec:concl}

The main aim of this paper has been to develop methodology that assists in the classification of newly observed FRBs as repeaters or non-repeaters.  Repeating  FRBs are of scientific interest and any such classification has the potential to increase the number of confirmed repeating FRBs at a more rapid pace. For example, by identifying repeater candidates from the first observation with a certain repeater probability, one could rank potential repeating FRBs for follow up. Alternatively, repeater candidates identified through our method may improve the efficiency of follow-up searches for repeat bursts in archival data.

The method proposed here is based on the widely used statistical model for studying the dependence of binary response variables on independent variables, known as logistic regression. Our logistic regression model is modified to account for the marked imbalance in the data (i.e., many fewer repeaters than non-repeaters). The adjustment of the method relies on a tuning parameter that represents the proportion of repeaters, $\tau$, in the whole population of FRBs.

Given the cross disciplinary nature of the methodology, we have prepared takeaways according to the main interest of the audience:

\paragraph{Astronomy Takeaways}

\begin{itemize}
    \item  One of the first takeaways is efficiency. Compared with a deep-learning approach or other machine learning classification techniques, logistic regression requires a smaller volume of data. 
    
    \item The performance of the model is promising for finding potential repeaters. For the Gold sample, the model correctly identifies 70\%  of the test set FRBs,  not previously used for training, as repeaters.

    \item While logistic regression is well-established in other fields, its application in astronomy, particularly for rare events, is relatively novel. We have introduced this approach to the astronomical literature and hope it can be successfully applied to other astronomy datasets that have binary responses. Moreover, we emphasize  the introduction of a parameter that could help interpret the true population fraction in any study containing an imbalanced data set. We encourage the application of our method to other datasets where a similar pattern is present.

\end{itemize}

\paragraph{Statistical Takeaways}

\begin{itemize}
    \item {The FRB data present interesting challenges for the statistician. The uncertain labeling of one-off FRBs is an open problem.} One cannot be sure that a one-off source will never repeat in the future. This uncertainty in labeling is an unusual issue in statistical applications of logistic regression. This implies that the model we propose should not be interpreted as a classifier. Rather, an astronomer interested in repeating FRBs will want to use our model to select and prioritize monitoring of one-off sources that are more likely to repeat. In other applications, where both cases and controls are unambiguously labeled, one can use a model like ours for classification.
    
    \item {The imbalance of the sample can be accounted for, with caveats.} The number of one-off FRBs vastly outnumbers the number of repeating FRBs in this application. This issue of imbalanced data is not unusual in statistical applications of logistic regression, and we have referenced works that tackle this problem. However, what is unusual in the application of FRBs is that we do not know the true, underlying proportion of repeating FRBs in the population. We have addressed this issue by allowing the proportion of repeaters in the population of FRBs, $\tau$, to take on various values, and by assessing performance metrics to settle on a range of probable values. We found values of $\tau$ in the range $0.5$ to $0.6$ to perform well, and noted that the results do not change significantly within this range. Thus, we presented the analysis when $\tau=0.55$. Our numerical experiments suggest that the model is most accurate when $\tau$ is around 60\%. However, we hesitate to conclude that this qualifies as strong statistical evidence in favor of this being the {\it true} proportion, since the tuning parameter is not estimable from the data. 
    
    \item {The dependence between observations produced by a repeating source stills needs to be taken into account.} A one-off source generates a single burst while a repeater generates one to several sub-bursts. In this paper, the information from multiple bursts is summarized by  the mean.  This strategy does not incorporate all the information contained in multiple bursts. Moreover, some repeating FRBs may have burst only twice, while others may have burst tens or even hundreds of times. Thus, some repeaters have more data available, but this is not considered in the training.  In other words, the amount of information provided by each repeater is not the same, a fact that the current model does not take into account. Moreover, sub-bursts cannot be treated as independent observations since they have single origin. We are currently exploring the use of a mixed effects weighted logistic regression model that will automatically integrate the information from all bursts that have the same source. The results will be communicated in a future paper.

\end{itemize}

\section{Acknowledgments}
This work was supported by a Collaborative Research Team grant to G.M.E. from the Canadian Statistical Sciences Institute (CANSSI), which is supported by Natural Sciences and Engineering Research Council of Canada (NSERC). We thank Amanda Cook for providing specifics about the CHIME/FRB collaboration, helpful comments, and minor editing of the paper. The Dunlap Institute is funded through an endowment established by the David Dunlap family and the University of Toronto. B.M.G. acknowledges the support of the NSERC through grant RGPIN-2022-03163, and of the Canada Research Chairs program. Z.P. was a Dunlap Fellow and is supported by an NWO Veni fellowship (VI.Veni.222.295). G.M.E also acknowledges the support of NSERC through Discovery Grant RGPIN-2020-04554. K.W.M. holds the Adam J. Burgasser Chair in Astrophysics.

\section{Author Contributions}

AHM performed all of the analysis and wrote all of the code for this project. AHM also made all figures and wrote the first and subsequent drafts of the paper. RVC, GME, and DCS co-supervised AHM and helped write and edit significant portions of the paper.

\nocite{*}

\bibliography{logpaper}{}

\begin{thebibliography}{}
\expandafter\ifx\csname natexlab\endcsname\relax\def\natexlab#1{#1}\fi
\providecommand{\url}[1]{\href{#1}{#1}}
\providecommand{\dodoi}[1]{doi:~\href{http://doi.org/#1}{\nolinkurl{#1}}}
\providecommand{\doeprint}[1]{\href{http://ascl.net/#1}{\nolinkurl{http://ascl.net/#1}}}
\providecommand{\doarXiv}[1]{\href{https://arxiv.org/abs/#1}{\nolinkurl{https://arxiv.org/abs/#1}}}

\bibitem[{Amiri {et~al.}(2018)Amiri, Bandura, Berger, Bhardwaj, Boyce, Boyle, Brar, Burhanpurkar, Chawla, Chowdhury, {et~al.}}]{amiri2018chime}
Amiri, M., Bandura, K., Berger, P., {et~al.} 2018, The Astrophysical Journal, 863, 48

\bibitem[{Amiri {et~al.}(2022)Amiri, Bandura, Boskovic, Chen, Cliche, Deng, Denman, Dobbs, Fandino, Foreman, {et~al.}}]{amiri2022overview}
Amiri, M., Bandura, K., Boskovic, A., {et~al.} 2022, The Astrophysical Journal Supplement Series, 261, 29

\bibitem[{Bailes(2022)}]{frbreview2022}
Bailes, M. 2022, Science, 378, eabj3043, \dodoi{10.1126/science.abj3043}

\bibitem[{Bhardwaj {et~al.}(2021)Bhardwaj, Gaensler, Kaspi, Landecker, Mckinven, Michilli, Pleunis, Tendulkar, Andersen, Boyle, Cassanelli, Chawla, Cook, Dobbs, Fonseca, Kaczmarek, Leung, Masui, Mnchmeyer, Ng, Rafiei-Ravandi, Scholz, Shin, Smith, Stairs, \& Zwaniga}]{Bhardwaj_2021}
Bhardwaj, M., Gaensler, B.~M., Kaspi, V.~M., {et~al.} 2021, The Astrophysical Journal Letters, 910, L18, \dodoi{10.3847/2041-8213/abeaa6}

\bibitem[{Chatterjee(2021)}]{shami2021}
Chatterjee, S. 2021, Astronomy \& Geophysics, 62, 1.29, \dodoi{10.1093/astrogeo/atab043}

\bibitem[{Chen {et~al.}(2021)Chen, Hashimoto, Goto, Kim, Santos, On, Lu, \& Hsiao}]{chen2021}
Chen, B.~H., Hashimoto, T., Goto, T., {et~al.} 2021, Monthly Notices of the Royal Astronomical Society, 509, 1227, \dodoi{10.1093/mnras/stab2994}

\bibitem[{{CHIME/FRB Collaboration} {et~al.}(2021){CHIME/FRB Collaboration}, Amiri, Andersen, Bandura, Berger, Bhardwaj, Boyce, Boyle, Brar, Breitman, Cassanelli, Chawla, Chen, Cliche, Cook, Cubranic, Curtin, Deng, Dobbs, Dong, Eadie, Fandino, Fonseca, Gaensler, Giri, Good, Halpern, Hill, Hinshaw, Josephy, Kaczmarek, Kader, Kania, Kaspi, Landecker, Lang, Leung, Li, Lin, Masui, Mckinven, Mena-Parra, Merryfield, Meyers, Michilli, Milutinovic, Mirhosseini, Münchmeyer, Naidu, Newburgh, Ng, Patel, Pen, Petroff, Pinsonneault-Marotte, Pleunis, Rafiei-Ravandi, Rahman, Ransom, Renard, Sanghavi, Scholz, Shaw, Shin, Siegel, Sikora, Singh, Smith, Stairs, Tan, Tendulkar, Vanderlinde, Wang, Wulf, \& Zwaniga}]{chime_2021}
{CHIME/FRB Collaboration}, Amiri, M., Andersen, B.~C., {et~al.} 2021, The Astrophysical Journal Supplement Series, 257, 59, \dodoi{10.3847/1538-4365/ac33ab}

\bibitem[{{CHIME/FRB Collaboration} {et~al.}(2023{\natexlab{a}}){CHIME/FRB Collaboration}, Andersen, Bandura, Bhardwaj, Boyle, Brar, Cassanelli, Chatterjee, Chawla, Cook, Curtin, Dobbs, Dong, Faber, Fandino, Fonseca, Gaensler, Giri, Herrera-Martin, Hill, Ibik, Josephy, Kaczmarek, Kader, Kaspi, Landecker, Lanman, Lazda, Leung, Lin, Masui, Mckinven, Mena-Parra, Meyers, Michilli, Ng, Pandhi, Pearlman, Pen, Petroff, Pleunis, Rafiei-Ravandi, Rahman, Ransom, Renard, Sand, Sanghavi, Scholz, Shah, Shin, Siegel, Smith, Stairs, Su, Tendulkar, Vanderlinde, Wang, Wulf, \& Zwaniga}]{chime2023}
{CHIME/FRB Collaboration}, Andersen, B.~C., Bandura, K., {et~al.} 2023{\natexlab{a}}, The Astrophysical Journal, 947, 83, \dodoi{10.3847/1538-4357/acc6c1}

\bibitem[{{CHIME/FRB Collaboration} {et~al.}(2023{\natexlab{b}}){CHIME/FRB Collaboration}, Amiri, Andersen, Bandura, Berger, Bhardwaj, Boyce, Boyle, Brar, Breitman, Cassanelli, Chawla, Chen, Cliche, Cook, Cubranic, Curtin, Deng, Dobbs, Dong, Eadie, Fandino, Fonseca, Gaensler, Giri, Good, Halpern, Hill, Hinshaw, Josephy, Kaczmarek, Kader, Kania, Kaspi, Landecker, Lang, Leung, Li, Lin, Masui, Mckinven, Mena-Parra, Merryfield, Meyers, Michilli, Milutinovic, Mirhosseini, Münchmeyer, Naidu, Newburgh, Ng, Patel, Pen, Petroff, Pinsonneault-Marotte, Pleunis, Rafiei-Ravandi, Rahman, Ransom, Renard, Sanghavi, Scholz, Shaw, Shin, Siegel, Sikora, Singh, Smith, Stairs, Tan, Tendulkar, Vanderlinde, Wang, Wulf, \& Zwaniga}]{err_chime_2023}
{CHIME/FRB Collaboration}, Amiri, M., Andersen, B.~C., {et~al.} 2023{\natexlab{b}}, The Astrophysical Journal Supplement Series, 264, 53, \dodoi{10.3847/1538-4365/acb54c}

\bibitem[{Connor \& van Leeuwen(2018)}]{Connor_2018}
Connor, L., \& van Leeuwen, J. 2018, The Astronomical Journal, 156, 256, \dodoi{10.3847/1538-3881/aae649}

\bibitem[{{Cordes} \& {Lazio}(2002)}]{cordeslazio2002}
{Cordes}, J.~M., \& {Lazio}, T.~J.~W. 2002, arXiv e-prints, astro, \dodoi{10.48550/arXiv.astro-ph/0207156}

\bibitem[{Cormack(1971)}]{cormack1971}
Cormack, R.~M. 1971, Journal of the Royal Statistical Society: Series A (General), 134, 321, \dodoi{https://doi.org/10.2307/2344237}

\bibitem[{Craiu \& Sun(2008)}]{craiu2008choosing}
Craiu, R.~V., \& Sun, L. 2008, Statistica Sinica, 861

\bibitem[{Cui {et~al.}(2020)Cui, Zhang, Wang, Zhang, Li, Peng, Zhu, Wang, Strom, Ye, Wang, \& Yang}]{cui2020}
Cui, X.-H., Zhang, C.-M., Wang, S.-Q., {et~al.} 2020, Monthly Notices of the Royal Astronomical Society, 500, 3275, \dodoi{10.1093/mnras/staa3351}

\bibitem[{{Fonseca} {et~al.}(2024){Fonseca}, {Pleunis}, {Breitman}, {Sand}, {Kharel}, {Boyle}, {Brar}, {Giri}, {Kaspi}, {Masui}, {Meyers}, {Patel}, {Scholz}, \& {Smith}}]{Fonseca2024ApJS}
{Fonseca}, E., {Pleunis}, Z., {Breitman}, D., {et~al.} 2024, \apjs, 271, 49, \dodoi{10.3847/1538-4365/ad27d6}

\bibitem[{Guo \& Wei(2022)}]{Guo_2022}
Guo, H.-Y., \& Wei, H. 2022, Journal of Cosmology and Astroparticle Physics, 2022, 010, \dodoi{10.1088/1475-7516/2022/07/010}

\bibitem[{He \& Garcia(2009)}]{imbalance2009}
He, H., \& Garcia, E.~A. 2009, IEEE Transactions on Knowledge and Data Engineering, 21, 1263, \dodoi{10.1109/TKDE.2008.239}

\bibitem[{Holland \& Welsch(1977)}]{robust1977}
Holland, P.~W., \& Welsch, R.~E. 1977, Communications in Statistics - Theory and Methods, 6, 813, \dodoi{10.1080/03610927708827533}

\bibitem[{Hung {et~al.}(2018)Hung, Jou, \& Huang}]{hung2018robust}
Hung, H., Jou, Z.-Y., \& Huang, S.-Y. 2018, Biometrics, 74, 145

\bibitem[{James(2023)}]{james2023modelling}
James, C.~W. 2023, Modelling repetition in zDM: a single population of repeating fast radio bursts can explain CHIME data.
\newblock \doarXiv{2306.17403}

\bibitem[{Kim \& Hwang(2022)}]{empiricalsamp2022}
Kim, M., \& Hwang, K.-B. 2022, PloS one, \dodoi{10.1371/journal.pone.0271260}

\bibitem[{King \& Zeng(2001{\natexlab{a}})}]{king2001}
King, G., \& Zeng, L. 2001{\natexlab{a}}, International Organization, 55, 693–715, \dodoi{10.1162/00208180152507597}

\bibitem[{King \& Zeng(2001{\natexlab{b}})}]{kingzeng2001}
---. 2001{\natexlab{b}}, Political Analysis, 9, 137–163, \dodoi{10.1093/oxfordjournals.pan.a004868}

\bibitem[{Kotsiantis {et~al.}(2006)Kotsiantis, Zaharakis, \& Pintelas}]{Kotsiantis2006}
Kotsiantis, S.~B., Zaharakis, I.~D., \& Pintelas, P.~E. 2006, Artificial Intelligence Review, 26, 159–190, \dodoi{https://doi.org/10.1007/s10462-007-9052-3}

\bibitem[{{Lin} \& {Sang}(2021)}]{Lin2021}
{Lin}, H.-N., \& {Sang}, Y. 2021, Chinese Physics C, 45, 125101, \dodoi{10.1088/1674-1137/ac2660}

\bibitem[{{Lingam} \& {Loeb}(2017)}]{lingam2017}
{Lingam}, M., \& {Loeb}, A. 2017, \apjl, 837, L23, \dodoi{10.3847/2041-8213/aa633e}

\bibitem[{Lorimer {et~al.}(2007)Lorimer, Bailes, McLaughlin, Narkevic, \& Crawford}]{lorimer2007}
Lorimer, D.~R., Bailes, M., McLaughlin, M.~A., Narkevic, D.~J., \& Crawford, F. 2007, Science, 318, 777, \dodoi{10.1126/science.1147532}

\bibitem[{Luo {et~al.}(2022)Luo, Zhu-Ge, \& Zhang}]{luo2022}
Luo, J.-W., Zhu-Ge, J.-M., \& Zhang, B. 2022, Monthly Notices of the Royal Astronomical Society, 518, 1629, \dodoi{10.1093/mnras/stac3206}

\bibitem[{Luque {et~al.}(2019)Luque, Carrasco, Martín, \& {de las Heras}}]{Luque2019}
Luque, A., Carrasco, A., Martín, A., \& {de las Heras}, A. 2019, Pattern Recognition, 91, 216, \dodoi{https://doi.org/10.1016/j.patcog.2019.02.023}

\bibitem[{Maalouf \& Siddiqi(2014)}]{maalouf2014}
Maalouf, M., \& Siddiqi, M. 2014, Knowledge-Based Systems, 59, 142, \dodoi{https://doi.org/10.1016/j.knosys.2014.01.012}

\bibitem[{Maalouf \& Trafalis(2011)}]{maalouf2011}
Maalouf, M., \& Trafalis, T.~B. 2011, Computational Statistics \& Data Analysis, 55, 168, \dodoi{https://doi.org/10.1016/j.csda.2010.06.014}

\bibitem[{Manski \& Lerman(1977)}]{manski1977}
Manski, C.~F., \& Lerman, S.~R. 1977, Econometrica, 45, 1977.
\newblock \url{http://www.jstor.org/stable/1914121}

\bibitem[{McCullagh \& Nelder(1989)}]{mccullagh1989generalized}
McCullagh, P., \& Nelder, J. 1989, Generalized linear models (Chapman and Hall)

\bibitem[{McCulloch(1997)}]{McCulloch1997}
McCulloch, C.~E. 1997, Journal of the American Statistical Association, 92, 162.
\newblock \url{http://www.jstor.org/stable/2291460}

\bibitem[{Nagelkerke \& Fidler(2015)}]{nagelkerke2015estimating}
Nagelkerke, N., \& Fidler, V. 2015, PLoS ONE, 10, e0140718

\bibitem[{Peng {et~al.}(2002)Peng, Lee, \& Ingersoll}]{Chao2002}
Peng, C.-Y.~J., Lee, K.~L., \& Ingersoll, G.~M. 2002, The Journal of Educational Research, 96, 3.
\newblock \url{http://www.jstor.org/stable/27542407}

\bibitem[{{Petroff} {et~al.}(2022){Petroff}, {Hessels}, \& {Lorimer}}]{petrof2022}
{Petroff}, E., {Hessels}, J.~W.~T., \& {Lorimer}, D.~R. 2022, \aapr, 30, 2, \dodoi{10.1007/s00159-022-00139-w}

\bibitem[{Pleunis {et~al.}(2021)Pleunis, Good, Kaspi, Mckinven, Ransom, Scholz, Bandura, Bhardwaj, Boyle, Brar, Cassanelli, Chawla, Dong, Fonseca, Gaensler, Josephy, Kaczmarek, Leung, Lin, Masui, Mena-Parra, Michilli, Ng, Patel, Rafiei-Ravandi, Rahman, Sanghavi, Shin, Smith, Stairs, \& Tendulkar}]{Pleunis_2021}
Pleunis, Z., Good, D.~C., Kaspi, V.~M., {et~al.} 2021, The Astrophysical Journal, 923, 1, \dodoi{10.3847/1538-4357/ac33ac}

\bibitem[{Sheather(2009)}]{Sheather2009}
Sheather, S.~J. 2009, Logistic Regression (New York, NY: Springer New York), 263--303, \dodoi{10.1007/978-0-387-09608-7_8}

\bibitem[{{Spitler} {et~al.}(2016){Spitler}, {Scholz}, {Hessels}, {Bogdanov}, {Brazier}, {Camilo}, {Chatterjee}, {Cordes}, {Crawford}, {Deneva}, {Ferdman}, {Freire}, {Kaspi}, {Lazarus}, {Lynch}, {Madsen}, {McLaughlin}, {Patel}, {Ransom}, {Seymour}, {Stairs}, {Stappers}, {van Leeuwen}, \& {Zhu}}]{Spitler2016}
{Spitler}, L.~G., {Scholz}, P., {Hessels}, J.~W.~T., {et~al.} 2016, \nat, 531, 202, \dodoi{10.1038/nature17168}

\bibitem[{Tomz {et~al.}(2003)Tomz, King, \& Zeng}]{King2003}
Tomz, M., King, G., \& Zeng, L. 2003, Journal of Statistical Software, 8, 1–27, \dodoi{10.18637/jss.v008.i02}

\bibitem[{van~den Goorbergh {et~al.}(2022)van~den Goorbergh, van Smeden, Timmerman, \& Van~Calster}]{goorbergh2022}
van~den Goorbergh, R., van Smeden, M., Timmerman, D., \& Van~Calster, B. 2022, Journal of the American Medical Informatics Association, 29, 1525, \dodoi{10.1093/jamia/ocac093}

\bibitem[{Wagstaff {et~al.}(2016)Wagstaff, Tang, Thompson, Khudikyan, Wyngaard, Deller, Palaniswamy, Tingay, \& Wayth}]{Wagstaff_2016}
Wagstaff, K.~L., Tang, B., Thompson, D.~R., {et~al.} 2016, Publications of the Astronomical Society of the Pacific, 128, 084503, \dodoi{10.1088/1538-3873/128/966/084503}

\bibitem[{Yamasaki {et~al.}(2023)Yamasaki, Goto, Ling, \& Hashimoto}]{yamasaki2023}
Yamasaki, S., Goto, T., Ling, C.-T., \& Hashimoto, T. 2023, Monthly Notices of the Royal Astronomical Society, 527, 11158, \dodoi{10.1093/mnras/stad3844}

\bibitem[{Yang {et~al.}(2023)Yang, Zhang, Wang, \& Wu}]{yang2023}
Yang, X., Zhang, S.-B., Wang, J.-S., \& Wu, X.-F. 2023, Monthly Notices of the Royal Astronomical Society, 522, 4342, \dodoi{10.1093/mnras/stad1304}

\bibitem[{{Yao} {et~al.}(2017){Yao}, {Manchester}, \& {Wang}}]{yaomanchester2017}
{Yao}, J.~M., {Manchester}, R.~N., \& {Wang}, N. 2017, \apj, 835, 29, \dodoi{10.3847/1538-4357/835/1/29}

\bibitem[{{Zhao} {et~al.}(2023){Zhao}, {Wang}, {Zhang}, {Zhang}, \& {Zhang}}]{Zhao2023}
{Zhao}, Z.-W., {Wang}, L.-F., {Zhang}, J.-G., {Zhang}, J.-F., \& {Zhang}, X. 2023, \jcap, 2023, 022, \dodoi{10.1088/1475-7516/2023/04/022}

\bibitem[{Zhu-Ge {et~al.}(2022{\natexlab{a}})Zhu-Ge, Luo, \& Zhang}]{Zhu-Ge2023}
Zhu-Ge, J.-M., Luo, J.-W., \& Zhang, B. 2022{\natexlab{a}}, Monthly Notices of the Royal Astronomical Society, 519, 1823, \dodoi{10.1093/mnras/stac3599}

\bibitem[{Zhu-Ge {et~al.}(2022{\natexlab{b}})Zhu-Ge, Luo, \& Zhang}]{luob2022}
---. 2022{\natexlab{b}}, Monthly Notices of the Royal Astronomical Society, 519, 1823, \dodoi{10.1093/mnras/stac3599}

\end{thebibliography}
\bibliographystyle{aasjournal}



\end{document}